\documentclass[12pt,preprint]{aastex}

\newcommand{\msun}{{\rm M_\odot}}

\shorttitle{Aspects of microlensing: Magnification distributions}
\shortauthors{Schechter, Wambsganss \& Lewis}

\begin{document}


\title{Qualitative aspects of quasar microlensing with two
mass components: magnification patterns and probability distributions}


\author{Paul L. Schechter}
\affil{Department of Physics, Massachusetts Institute of Technology, 
77 Massachusetts Avenue, Cambridge, MA 02139}
\email{schech@achernar.mit.edu}

\author{Joachim Wambsganss}
\affil{Universit\"{a}t Potsdam, Institut f\"{u}r Physik, 
Am Neuen Palais 10, 14467 Potsdam, Germany}
\email{jkw@astro.physik.uni-potsdam.de}

\and

\author{Geraint F. Lewis}
\affil{School of Physics, A29, University of Sydney, 
Sydney, NSW 2006, Australia}
\email{gfl@physics.usyd.edu.au}



\begin{abstract}

It has been conjectured that the distribution of magnifications of a
point source microlensed by a randomly distributed population of 
intervening point masses is independent of its mass spectrum.  
We present {\it gedanken} experiments that cast doubt on this 
conjecture and numerical simulations that show it to be false.

\end{abstract}





\section{INTRODUCTION}\label{introduction}

Every  investigation of microlensing  at high  optical depth  that has
explored the effect of multiple microlens mass components  has led to
the  conclusion  that the  magnification  probability distribution  is
independent of  the spectrum of  microlens masses.  The recent
effort by \citet{2001MNRAS.320...21W} is typical.  While it was
not their principal result, they comment in passing
\begin{quotation}
``...   we  confirm   the  finding  of  \citet{1992ApJ...386...19W}  and
\citet{1995MNRAS.276..103L}  that  the  magnification distribution  is
independent of the mass function.''
\end{quotation}
This conjecture has important consequences regarding the more general
applicability of microlensing studies that are limited to a single 
mass component.  While galaxies have stars with a range of masses,
restricting to a single component makes analytic calculations more tractable
\citep[e.g.][]{
1986MNRAS.223..113P,1987ApJ...319....9S,1997ApJ...489..508K}
and greatly decreases the number of cases that must be simulated numerically
\citep[e.g.][]{1992ApJ...386...19W,1995MNRAS.276..103L,2001MNRAS.320...21W}. 
If true, the conjecture simplifies things considerably.

Both theoretical and experimental lines of evidence lead to this
conclusion, which has struck many investigators as obvious.  On
the experimental side, simulations like those carried out by
\citet{2001MNRAS.320...21W} and their predecessors produce
magnification histograms for different mass distributions that appear
to be indistinguishable for fixed surface mass density and shear.

On  the   theoretical  side,  the  high  magnification   tail  of  the
magnification   probability  distribution   has  been   shown   to  be
independent       of      the       microlens       mass      spectrum
\citep{1987ApJ...319....9S}.    Moreover,  \citet{1992A&A...258..591W}
showed that the average  number of positive parity microimages depends
only upon  the surface mass density (or equivalently the convergence)
and the shear.   Since the scale free nature  of gravity requires that
the magnification  probability distribution for a point  source be the
same for  microlensing by a single  mass of any size,  it would appear
strange if a mixture of two masses (at constant convergence and shear)
produced a different magnification probability distribution.

There is, however, at least one argument against this apparently
obvious conclusion, which we detail in $\S$ 2 below.  It suggests that
the magnification probability distribution {\it does} depend upon the
mass spectrum.  The argument suggests that the dependence would show
up in a highly magnified negative parity macroimage -- typically one
of a close pair of images in a quadruply imaged quasar like PG1115+080.

We have carried out lensing simulations of such an image (at constant
convergence and shear) for a variety of different cases.  In Figure 1
we show simulations with two populations of point masses.  The first
component is comprised of $1.000 \msun$ objects referred to hereafter
as ``micro-lenses.''  The second component is comprised of $0.005
\msun$ objects referred to hereafter as ``nano-lenses.''  The
designations and mass scale are arbitrary but are intended to convey the
sense that the micro-lenses are very much smaller than the lensing
galaxy and that the nano-lenses are very much smaller than the micro-lenses.

The eight panels of Figure 1 show magnification histograms obtained by
varying the mass fractions in the micro-lensing component, with the
remaining fraction in the nano-lensing component.  For the sake of
comparison, we reproduce in each panel the result for a pure
micro-lensing component.  As the fraction contributed by micro-lenses
decreases to 20\% and 10\% the histogram broadens out and develops a
second peak.  But as it decreases further to 0\%, the magnification
distribution narrows and ends up looking like the 100\% case (modulo
finite source effects and sample variance).  Unless our simulations
are faulty, the conjecture is false.

In $\S$ 2 we put forward  a qualitative argument for the dependence of
the microlensing  probability distribution  on the mass  spectrum.  In
$\S$ 3 we  give details of the numerical  simulations that confirm the
effect.  In $\S$ 4 we offer a qualitative interpretation of our results.  In
$\S$ 5 we discuss some astrophysical consequences.

\section{AN ARGUMENT FOR THE DEPENDENCE OF THE MAGNIFICATION
PROBABILITY DISTRIBUTION ON THE MASS SPECTRUM}

In Figure 2a we show the magnification probability distribution for a
simulation of a negative parity macroimage with convergence $\kappa =
0.55$ and shear $\gamma = 0.55$.  In this simulation all of the mass
is in micro-lenses of a single mass.  In Figure 2b we again show
the magnification probability distribution for a simulation of a
negative parity macroimage with convergence $\kappa = 0.55$ and shear
$\gamma = 0.55$, but in this case 20\% of the mass is in micro-lenses
of a single mass and 80\% of the mass is in a smooth mass sheet.  The
two histograms look quite different, with the first showing a single
peak and the second being significantly broader and showing two
peaks.\footnote{The bi- and even tri-modality of magnification
histograms has frequently been noted
\citep{1992ApJ...386...30R,1992ApJ...386...19W,1995MNRAS.276..103L,2002ApJ...580..685S}. The peaks can be indexed by the number of ``extra'' positive parity
micro-images \citep{1992ApJ...386...30R,2003ApJ...583..575G}.  
The broadening of magnification histograms at 
intermediate magnifications has likewise known for some time
\citep{1994A&A...288...19S,2002ApJ...580..685S}.}

Now suppose that the smooth mass sheet of Figure 2b is divided into
randomly distributed point masses that are very much smaller than the
micro-lenses.  We then have a micro-lensing component with
$\kappa_{micro} = 0.11$ and a nano-lensing component with $\kappa_{\rm
nano} = 0.44$.  If the hypothesis that the magnification distribution
is independent of the mass spectrum were correct, the magnification
probability distribution would look the same as that of Figure
2a.\footnote{An anonymous referee has argued that breaking up the
smooth sheet into small clumps can only broaden the magnification
histogram and that the conjecture must therefore be incorrect.  This
is borne out by the simulations presented in the following section.}

Finally, suppose we take our source to be {\it extended} rather than
point-like.  In particular, we imagine our source is much larger than
the Einstein rings of our nano-lenses but much smaller than the
Einstein rings of our micro-lenses.  The nano-lenses should behave
like a smooth component and the magnification probability distribution
should look like Figure 2b.

Alternatively, we can compute the magnification probability
distribution for our extended source by taking the magnification map
for a point source and convolving it with the surface brightness
distribution of the extended source.  Such a convolution will
inevitably smooth the map out, increasing the values of low
magnification pixels and decreasing the values of high magnification
pixels.  If the conjecture were correct and the magnification
histogram for a point source and macro- and nano-lensing components
looked like Figure 2a, we would expect the magnification probability
distribution for an extended source to be {\it narrower}.

Our two alternative schemes for computing the magnification histogram
of an extended source lensed by a two component screen give different
histograms, in one case broader and in the other case narrower than
the histogram of Figure 2a.  The histogram cannot simultaneously
be both narrower and broader than that of Figure 2a.  There
is a bad link in one of the chains of argument, which we take to be the
assumption that the magnification histogram for a point source is
independent of the mass spectrum.

\section{MICROLENSING SIMULATIONS}\label{numerical}

The  particular  values  for  the  convergence,  shear,  and  relative
fractions in the two mass components used in the previous section were
chosen (guided by the results of  Granot et. al. 2003) to maximize the
difference  between   Figures  2a  and   2b.   We  have   carried  out
microlensing simulations  at the same  convergence and shear,  to look
for   differences  in  the   microlensing  histogram   with  different
proportions of components.

The simulations  were done  employing the inverse  ray-shooting method
\citep{1986A&A...166...36K,1987A&A...171...49S}    as   described   in
Wambsganss (1990,  1999).  We  used a square  receiving field  of side
length 20  Einstein radii  $R_E$ (of the  high mass  component $m_{\rm
micro}$)\footnote{$R_E$ is the Einstein radius and is the natural scale
length for  gravitational microlensing.  In  the source plane,  $R_E =
\sqrt{ (4  G M / c^2  ) (D_{os} D_{ls}  / D_{ol})}$, where $M$  is the
mass of the microlensing object,  and $D_{ij}$ is the angular diameter
distances between observer ($o$), lens ($l$) and source ($s$); $c$ and
$G$  are  the  velocity  of  light  and  the  gravitational  constant,
respectively.}.   This area  in the  source  plane was  covered by  25
million pixels ($5000^2$).  We  used values of $\kappa_{\mathrm tot} =
0.55$ and $\gamma = 0.55$ for surface mass density and external shear,
corresponding  to  an  (average)   magnification  of  $|\mu|  =  10.0$
(negative  parity).   The positions  of  the  lenses were  distributed
randomly in  a circle significantly  larger than the  shooting region.
The total  number of  rays per frame  was typically  about $n_{\mathrm
{rays}}  \approx 10^{10}$,  resulting in  over 200  rays per  pixel on
average (the shooting  region was larger than the  receiving region so
that a significant number of rays landed outside the latter).

We   performed   a   series   of  simulations   with   changing   mass
components.  For the  first series, we used two  mass components with a
mass ratio  of $m_{\rm micro}/m_{\rm  nano} = 200$.  For  specificity
we adopted $m_{\rm micro} = 1\msun$, appropriate to stars and $m_{\rm
nano} = 0.005 \msun$, as might apply to very massive planets.

We started with three cases: in the first
case, 100\% of the mass was in micro-lenses with mass $m_{\rm micro}$;
in the second 50\% of the micro-lenses were replaced with smooth
matter; in the third case 50\% of the matter was in nano-lenses with
mass $m_{\rm nano}$ rather than in smooth matter.

The magnification maps for these simulations are displayed in
Figure~3, with the left-hand panel presenting the full 20$R_E$,
whereas the right-hand panel focuses upon a 1$R_E$ part of the
map. For the smooth matter case (central panels) and the $m_{\rm
nano}$ scenario (lowest panels), the location of the $m_{\rm micro}$
objects are the same.

In comparing the panels, it is apparent that the smooth matter and
$m_{\rm nano}$ simulations possess similar large scale structure in
their magnification maps, structure which is somewhat different from the
case where all the mass is in $m_{\rm micro}$ objects.  On smaller
scales, however, the magnification patterns for the smooth mass and
$m_{\rm nano}$ cases are quite different, with the presence of the smaller
mass nano-lenses breaking up the magnification structure into smaller
scale caustics.

The magnification distributions for these simulations are presented in
Figure 4. As  discussed previously, the case where all  the mass is in
$m_{\rm micro}$ objects is unimodal,  with the smooth matter case being
bimodal . The case containing $m_{\rm nano}$ masses clearly differs from
the solely $m_{\rm micro}$ case,  also appearing bimodal and similar in
form to  the smooth matter  case, at odds with the conjecture.

An examination of the magnification maps in Figure 3 illuminates the
differences between the magnification distributions in Figure 4.
Compared with the smooth matter map, the map for 100\% $m_{\rm micro}$
has a higher density of caustics and fewer regions of demagnification
(light-grey).  These regions of the source plane produce no positive
parity images.  Crossing caustics produces extra positive parity
images and additional magnification.  These regions dominate the
magnification histogram.  In the smooth matter case, the magnification
map has been `opened up,' revealing more extended regions with no
positive parity image and enhancing the low magnification peak seen in
the magnification distribution.  On large scales the magnification
distribution for the 50\% $m_{\rm nano}$ case resembles that of the
smooth matter case, again with larger regions without positive
parity images.  Thus its magnification probability distribution looks
more like that of the smooth matter case than that of the 100\% $m_{\rm
micro}$ case.  Indeed the 50\% $m_{\rm nano}$ case is even broader
than the smooth case, due to the additional corrugation of the 
large scale map by the small scale lenses.

Further simulations were undertaken in an attempt to understand this
difference. Again, we started with 100\% micro-lenses, $m_{\rm
micro}$.  Then we put 1\% of the total mass in nano-lenses, $m_{\rm
nano}$, (re-)distributing them randomly over the lens plane.  We
increased the nano-lens mass fraction to 2\%, 5\%, 10\%, and then
proceeded in steps of 10\% to 90\%.  We ended symmetrically with 95\%,
98\%, 99\% and 100\% nano-lenses, for a total 17 different cases.  The
numbers of lenses ranged from 25,000 (for 100\% m$_{\rm micro}$) to
2,600,000 (for 100\% m$_{\rm nano}$).

A selection of the resulting two dimensional magnification patterns is
shown in Figure 5.  The top six panels show the full simulation, while
the bottom six panels show an expanded inset.  Particularly notable is
the similarity between the upper left panel (with all the mass in
micro-lenses) and the lower right panel (the blowup of the map when
all the mass is in nano-lenses).  Simulations of this series were used
to produce the histograms in Figure 1.

For the final series, we took 50\% of the mass to be in micro-lenses,
and 50\% in nano-lenses, but let the masses of the nano-lenses vary
with $m_{\rm nano}/m_{\rm micro} = 0.32, 0.10, 0.032, 0.01,$ and
$0.0032$.  As a bracketting cases we considered $m_{\rm nano}/m_{\rm
micro} = 1$ and the 50\% smooth case, corresponding to $m_{\rm
nano}/m_{\rm micro} \rightarrow 0$, making seven cases altogether.
The magnification patterns for all but the smooth case are displayed
in Figure 6.  The magnification distributions are seen in Figure 7.

This last series shows that the conjecture fails only gradually. The
presence of the second component becomes significant (for our simulation)
only when the nano-lens masses are one tenth those of the micro-lenses.
By the time the nano-lenses are one hundredth those of the micro-lenses,
the effect is as large as we can measure.  In hindsight this 
onset would have been more appreciable had we put 80\% of the mass
into nano-lenses (as in the third panel of Figure 1), but the qualitative
effects would have been the same.

Our lensing simulations have two limitations:
\begin{itemize}
\item With the limited size of $L = 20 R_E$, the magnification maps
exhibit features that are correlated on scales uncomfortably close to
the size of the simulation volume.  An ensemble of simulations with
the same parameters will exhibit differences due to sample variance.
We checked this for a few cases and found this sample variance to be
small compared to the observed differences needed to make our case.
Moreover, in the series of simulations described above, we kept the
positions of the stars fixed to the extent possible, so that we could
study the differential changes from one case to the next, and to hence
minimize the effects of sample variance.

\item  The finite  pixel size  corresponds to  a minimum  source size,
i.e. our results are not quite applicable to a point source.  However,
a size of $20 R_E/5000{\mathrm {pix}} = 0.004 R_E/{\mathrm {pix}}$, is
small  enough for  the effects  we want  to study  and  explore.  This
finite  size unavoidably  cuts off  the magnification  distribution at
very high magnifications $\mu$ and  leads to deviations from the power
law behavior, but the low and intermediate magnification region we are
interested in (see next section) is not strongly affected by that.

\end{itemize}

Despite the inevitable limited dynamic range for such simulations, we
have tried to choose parameters such that we can demonstrate the
effect most convincingly.

\section{INTERPRETATION} \label{discussion}

In the previous section we simulated cuts through the $\kappa_{\rm
nano}/\kappa_{\rm micro}, m_{\rm nano}/m_{\rm micro}$ plane at fixed
$\kappa_{tot}$ and $\gamma$.  Somewhat counter-intuitively, we find
that at fixed mass ratio $m_{\rm nano}/m_{\rm micro} = 1/200$ the
magnification probability histogram is broader for comparable amounts
of two very different masses than it is for a single component of
either one mass or the other.

The scale invariance of gravity demands that, for a point source, the
magnification histograms of single components of very different masses
should be identical.  But our experiments show that for two very
different mass components the magnification map looks very much like
that of the higher mass component immersed in a perfectly smooth
component.  Only on small scales are there differences.  This can be
seen in Figure 3.

Suppose one grants that the magnification probability distribution for
two very disparate components looks like that for a single component
and a smooth component.  The arguments set forth in
\citep{1992ApJ...386...30R}, Schechter and Wambsganss (2002) and
Granot et al. (2003) would come into play: the magnification histogram
tends to be broadest when the effective magnification computed from
the effective convergence and the effective shear is of order $|\mu|
\approx 3-4$.  Alternatively, the fluctuations are largest when the
number of extra positive parity images is roughly unity.  In such
cases one tends to get two peaks.

But the magnification probability distribution for two disparate
components is not {\it exactly} that of a single component and a
smooth component.  The low mass component produces additional
structure in the magnification map, further broadening the
magnification histogram, rounding off its peaks and filling in its
valleys.  There is evidence for this in Figures 2 and 3.

Once one substitutes the low mass component for a smooth component,
the number of extra positive parity images increases from roughly
unity to something significantly larger.  While this tends to round of
the two peaks, it does {\it not} narrow the magnification histogram.
The arguments of Schechter \& Wambsganss (2002) and Granot et
al. (2003) that the magnification histogram is broadest when there is,
on average, one extra positive parity image does not hold for two
disparate mass components.  The reason is that the extra positive
parity images cluster around the images produced by the single mass
component, breaking them into pieces but only slightly changing the
combined contribution to the flux.

\section{ASTROPHYSICAL CONSEQUENCES}

In gravitational lensing, magnifications depend upon second
derivatives (with respect to position) of the time delay function
\citep[e.g.][]{1986ApJ...310..568B}.  Deflections depend upon first
derivatives.  And time delays depend upon the function itself.  The
second derivatives are dimensionless, with the consequence that the
magnification of an image (unlike its deflection and time delay)
contains no information about the mass of intervening lens.

Image position fluctuations due to microlensing manifestly {\it do}
contain information about the mass scale of the intervening
microlenses \citep[][and references
therein]{1998ApJ...501..478L,2004A&A...416...19T}.  Moreover the
timescale over which brightness fluctuations occur likewise contains
information about the mass scale of the intervening microlenses
(and on the distribution of microlens masses)
if one knows the relative velocities of the microlenses and the source
\citep{2001MNRAS.320...21W}.  But the amplitude of those brightness
fluctuations is independent of mass scale.

In the present paper we consider the dependence of brightness
fluctuations not on mass scale, the first moment of the microlens mass
distribution function, but on higher order dimensionless moments of
that mass distribution.  We have demonstrated (through our simulations
using two mass components) that the magnification probability does
depend upon those higher moments.  We have not, however, explored the
full range of astrophysically interesting mass distributions.

We have chosen to explore in detail the specific case of two mass
components with $\kappa_{tot} = \gamma = 0.55$, appropritate to
one of two images in a highly magnified pair, as in the case of
PG1115+080 \citep{1981ApJ...244..723Y} or SDSS0924+0219
\citep{2003AJ....126..666I}.  The argument of Section 2 led us to
believe that the effects of using two mass components rather than a
single mass component would be appreciable in this case.  But what
about other values of the convergence and shear?  How much does the
mass spectrum matter for images of a quasar which are not
saddle-points, or not highly magnified?

A thorough treatment of this question would explore a substantial
fraction of the $\kappa,\gamma$ plane, and would quantify with some
statistic the differences between a single mass component and a range
of masses.  Such a treatment lies beyond the scope of the present
paper.

The fact that most previous investigators have failed to detect the
effects of a range of microlens masses would argue that to first order
such effects can be ignored.  Even in the present case, where the
convergence and shear have been chosen to maximize the effects, they
are not large.  Most mass distributions tend to put most of the mass
at one or the other end of the mass distribution.  The present
simulations would seem to indicate that only if appreciable mass
fractions are in components that differ by more than a factor of ten
in mass will the effects of a range of masses be substantial.

\section{ACKNOWLEDGMENTS}
We gratefully acknowledge helpful conversations with Arlie Petters.
GFL and JW thank M.I.T.'s Center for Space Resarch for its hospitality
during a collaboratory visit during which much of this work was
undertaken. GFL acknowledges losing a bottle of whiskey to PLS for
backing the wrong horse.  PLS gratefully acknowledges the whiskey.
The work was supported in part by US NSF grant AST-020601.

\clearpage
\begin{figure}
\epsscale{1.0}
\plotone{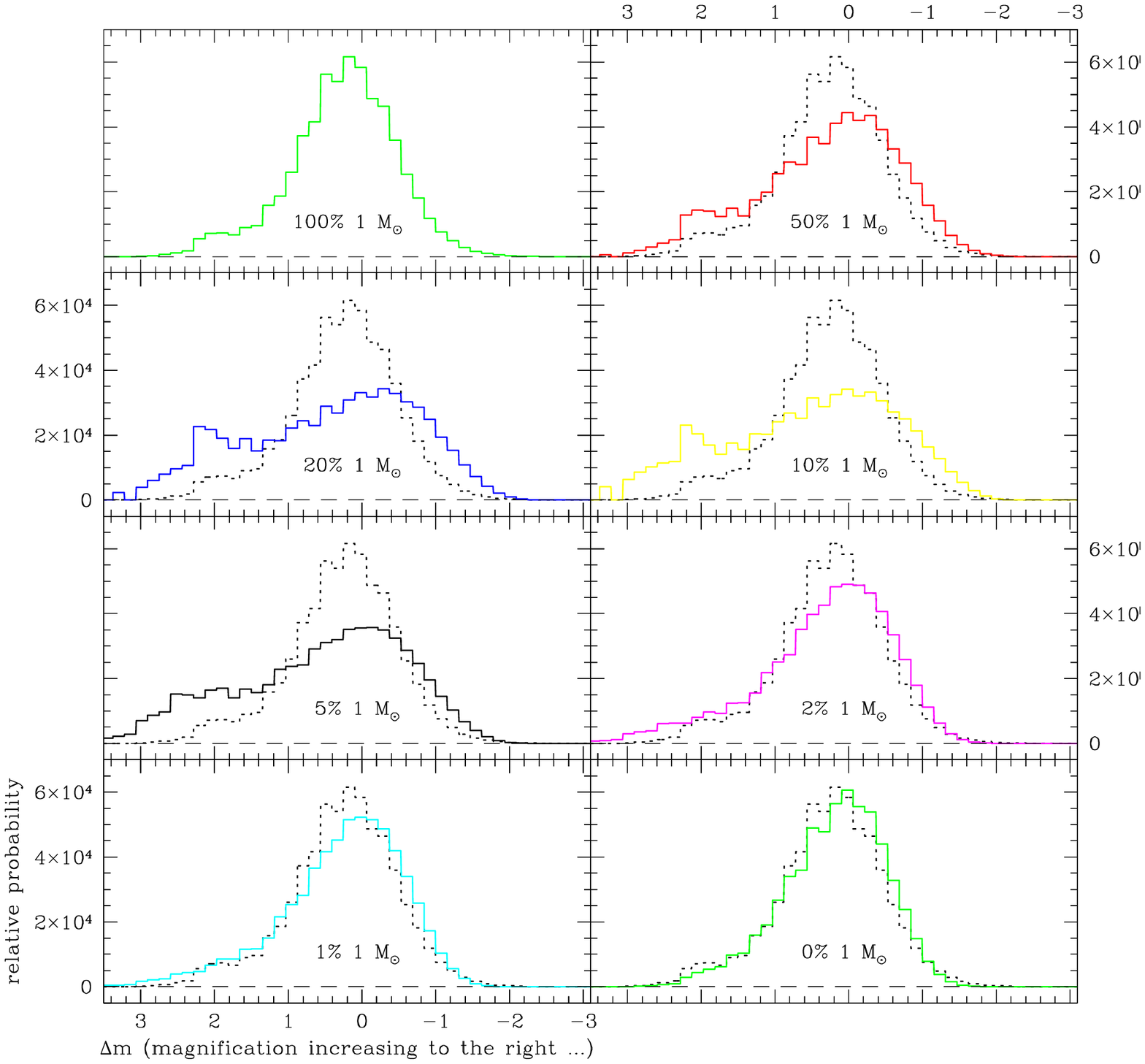}
\caption{Magnification       probability       distributions       for
$\kappa_{tot}=\gamma=0.55$. The  percentage in each  panel denoted the
fraction of $\kappa_{tot}$ composed of 1$\msun$ objects, the remainder
being in  0.005$\msun$ masses.  The dotted-line in  each panel  is the
magnification probability  distribution for the case  where the entire
microlensing population is comprised of 1$\msun$ object (presented in the
upper left-hand panel).\label{fig5}}
\end{figure}

\begin{figure}
\epsscale{1.0}
\plottwo{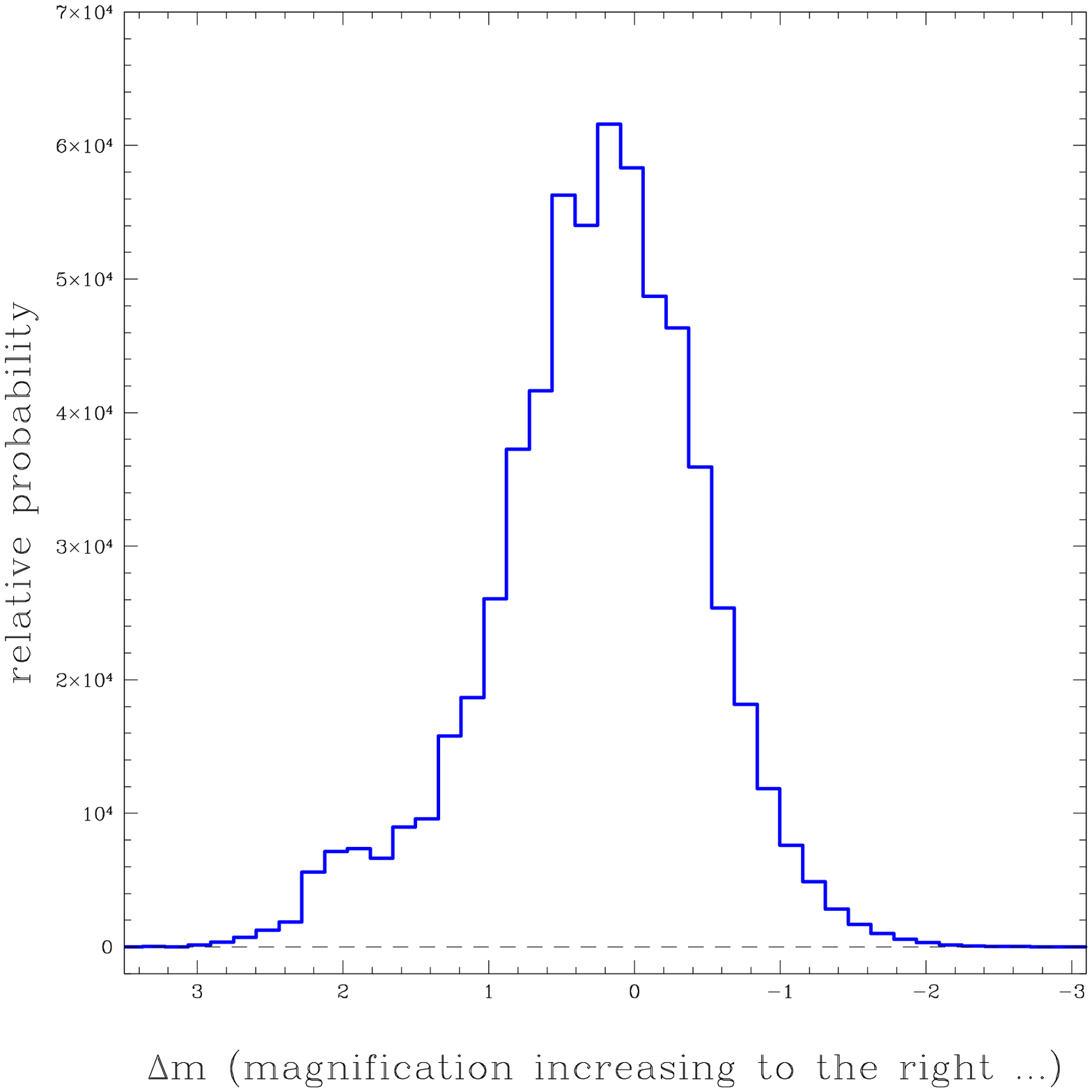}{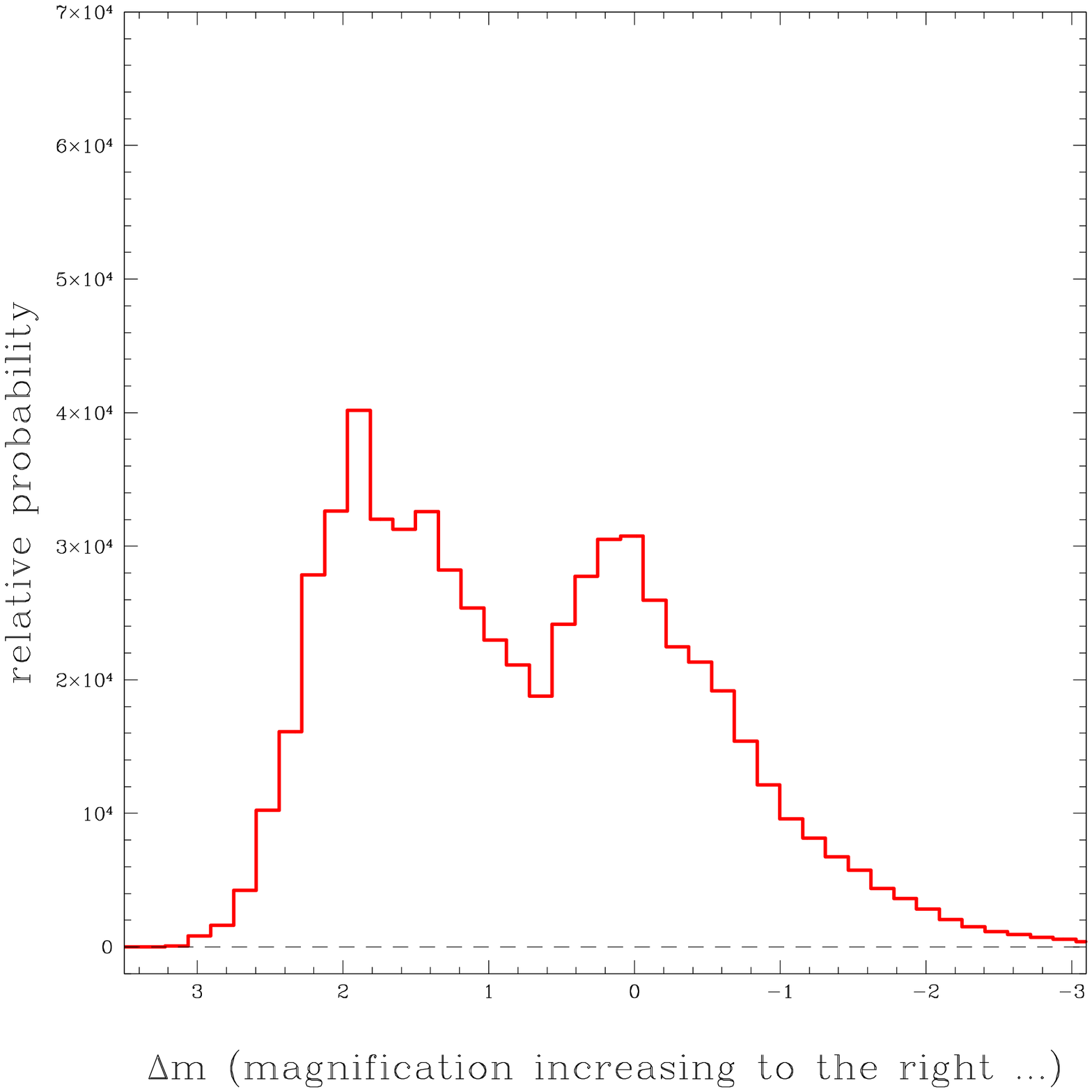}
\caption{Magnification       probability       distributions       for
$\kappa_{tot}=\gamma=0.55$.   On the left 
100\% of the matter is in $m_{\rm micro} = 1\msun$ objects.  On the right
20\% of the matter is in $m_{\rm micro} = 1\msun$ objects and 80\% is
in a smooth mass sheet.
\label{figx}}
\end{figure}

\clearpage
\begin{figure}
\epsscale{.80}
\plotone{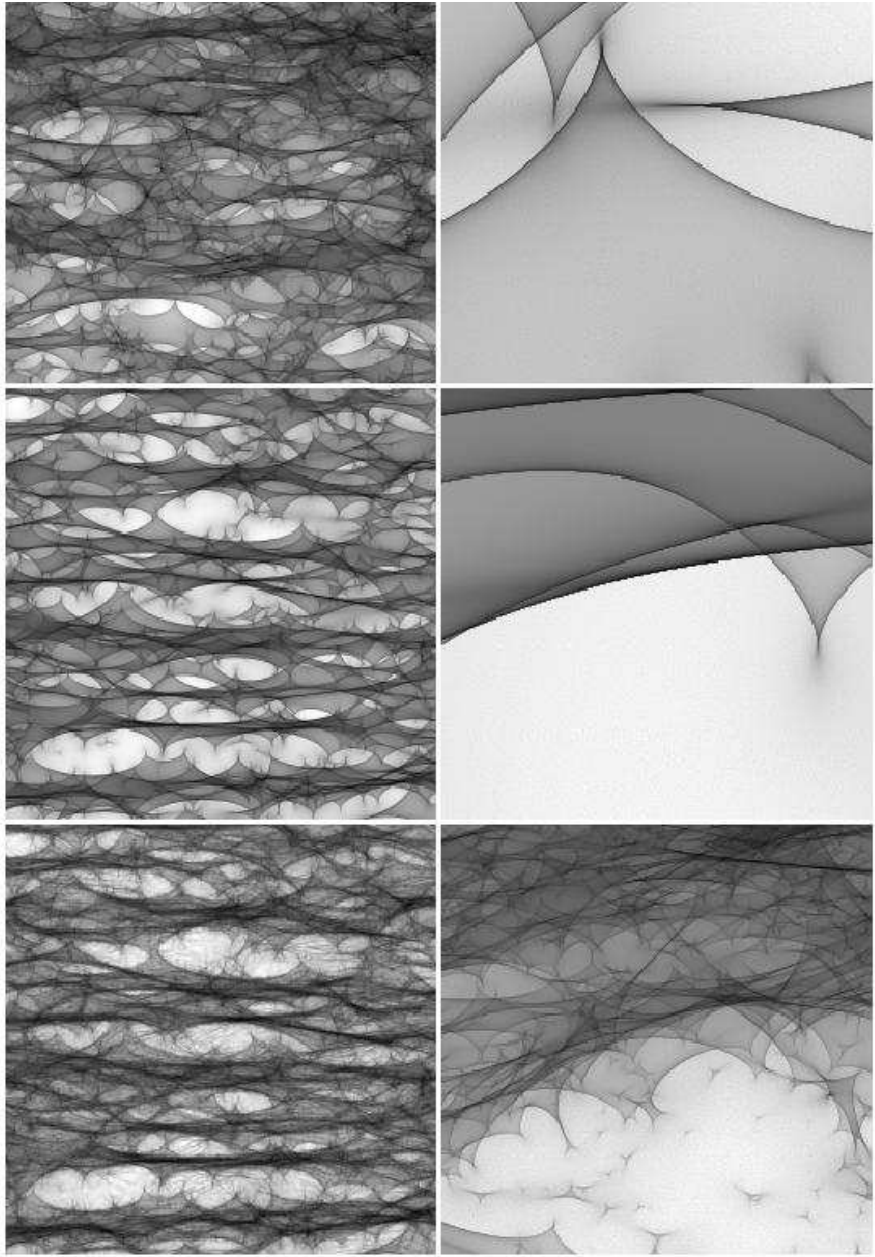}
\caption{Magnification      maps      for      the      case      with
$\kappa_{tot}=\gamma=0.55$.  The left-hand  figures represent a region
20 $R_E$ (for a solar  mass star) on a side. The right-hand  panel is a zoom
of the  lower left-hand  1 $R_E$. 
The top  row maps were  constructed with
$\kappa_{tot}$  all in  1$\msun$  masses, 
whereas  the central  panels
consist of  a microlensing population  with 50\% of  $\kappa_{tot}$ in
1$\msun$ objects and the rest in continuous matter.
The lower
panels are for 50\% of $\kappa_{tot}$ in 1$\msun$ objects and 
the remainder in 0.005$\msun$ masses. 
\label{fig1}}
\end{figure}

\clearpage
\begin{figure}
\epsscale{1.0}
\plotone{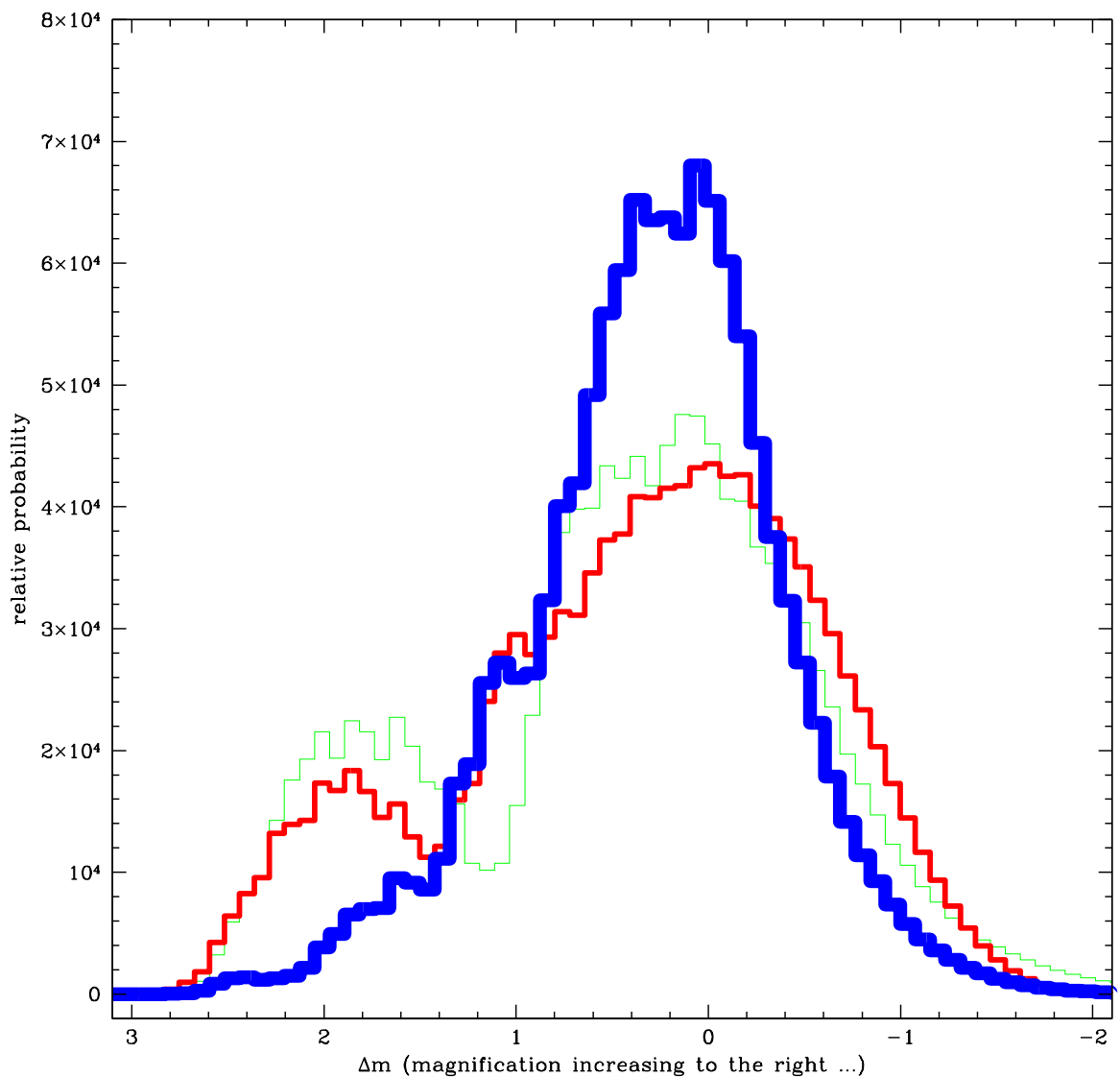}
\caption{The     magnification    probability     distributions    for
$\kappa_{tot}=\gamma=0.55$.  The thick curve  presents the  case where
all  the mass  is in  1$\msun$ objects,  whereas the  medium thickness
curve is  when 50\% of $\kappa_{tot}$  is in 1$\msun$  objects and the
remainder is in continuous matter. For the thin curve, this continuous
matter  contribution has  been  replaced  by objects  with  a mass  of
0.005$\msun$.\label{fig2}}
\end{figure}

\clearpage
\begin{figure}
\epsscale{.80}
\plotone{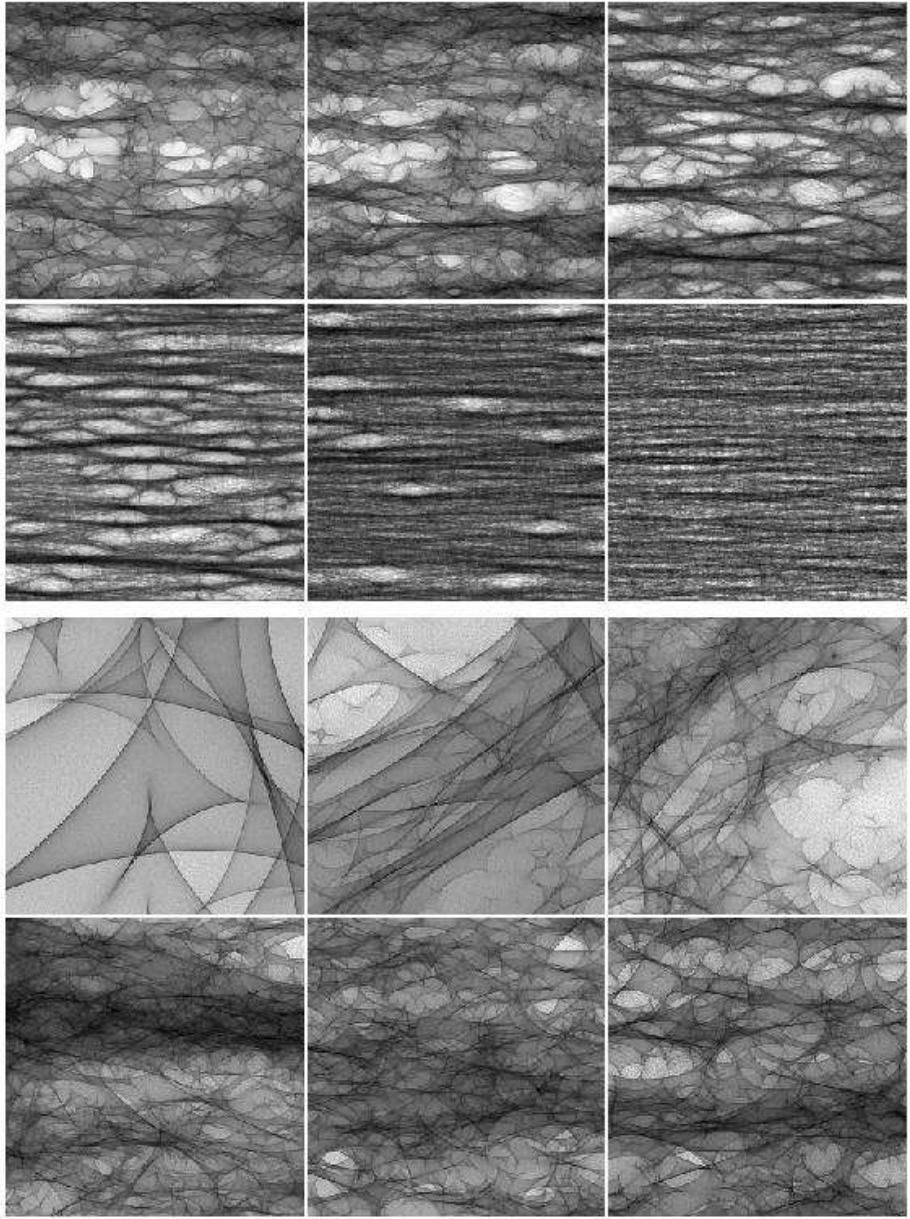}
\caption{Magnification  maps for the  case $\kappa_{tot}=\gamma=0.55$.
In the top six panels,
the proportion of microlenses with a particular mass is
changed   such  that  (from   left-to-right  and   top-to-bottom)  the
percentage of $\kappa_{tot}$ in 0.005$\msun$ objects is 0\%, 20\%, 
60\%, 90\%, 98\% \& 100\%; the remainder
of  $\kappa_{tot}$   is  in  1$\msun$   masses  in  each   case.  Each
magnification panel is $L = 20 R_E$ in extent.
The size panels at the bottom show zooms of the lower left
corners of the same sequence, respectively, sidelengths 
here are L = 1 $R_E$ (defined for a 1$\msun$-lens).
\label{fig3}}
\end{figure}

\clearpage
\begin{figure}
\epsscale{.80}
\plotone{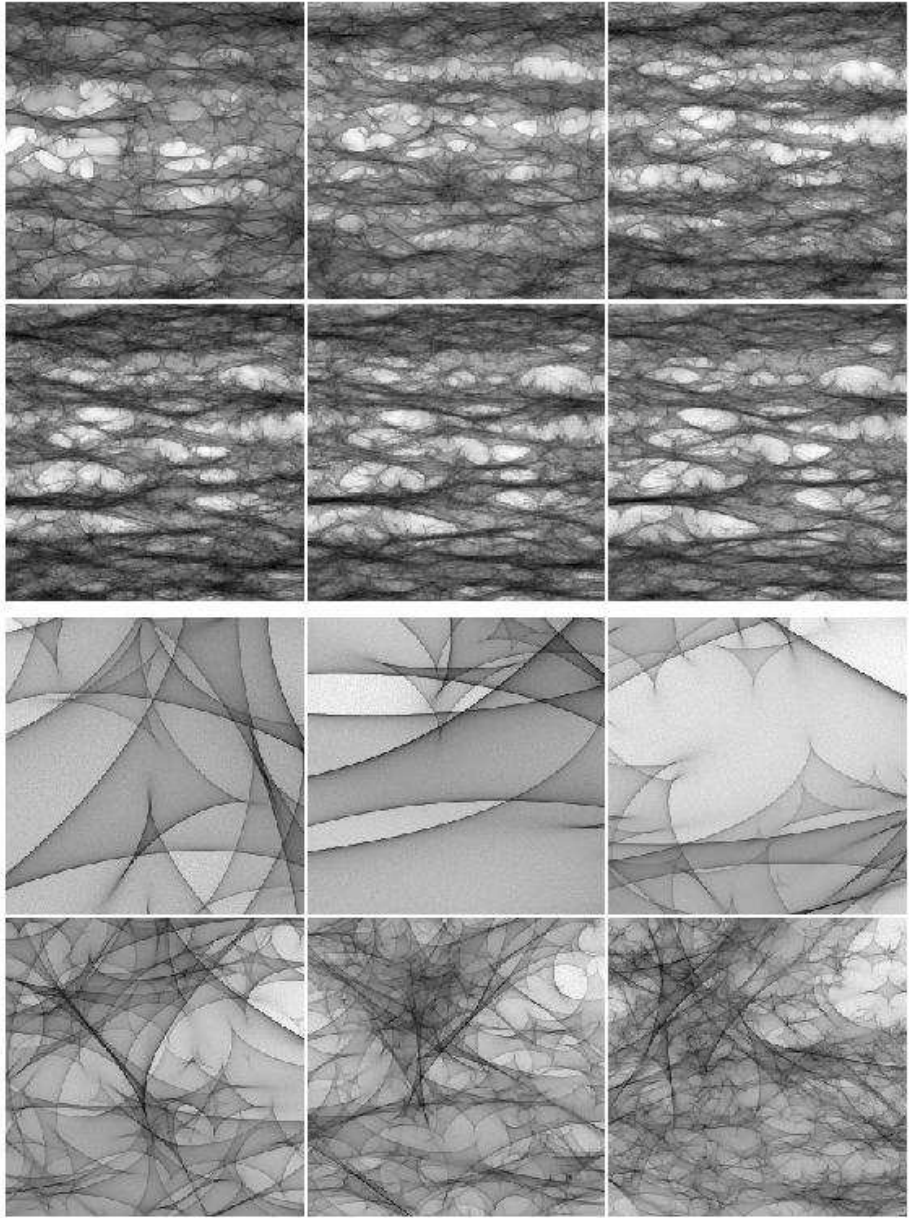}
\caption{The  top six panels  present the  magnification maps  for the
case  $\kappa_{tot}=\gamma=0.55$. In each,  50\% of  $\kappa_{tot}$ is
comprised of 1$\msun$ objects, while the remainder is composed of (top
row,  from  left to  right)  1$\msun$, $\sqrt{0.1}\msun$,  $0.1\msun$,
(second row) $\sqrt{0.01}\msun$, $0.01\msun$ and $\sqrt{0.001}\msun$.
Each panel is 20 $R_E$ in extent. The lower six panels present a zoom
of the lower left-hand corner of the same magnification maps (extend 
is L = 1 $R_E$).
\label{fig4}}
\end{figure}

\clearpage
\begin{figure}
\epsscale{.90}
\plotone{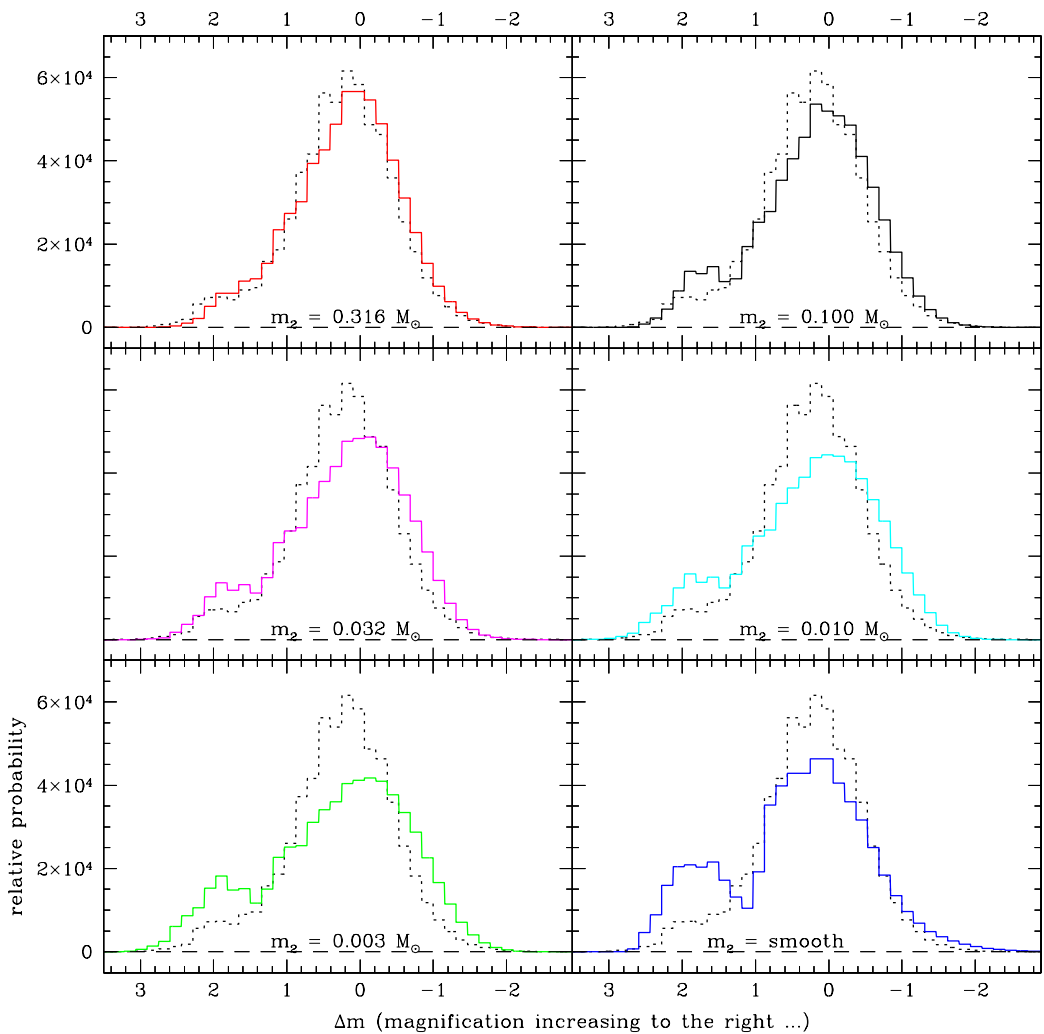} 
\caption{The six panels  present the  magnification distributions
for a scenario with 
$\kappa_{tot}=\gamma=0.55$, 
and the matter being split between evenly 
two mass components (50\% $m_{\rm micro}$, 50\% $m_{\rm nano}$).
The mass ratios are $m_{\rm micro}/m_{\rm nano} = 
0.316, 0.100, 0.032, 0.01, 0.003$ for the first 
five panels, and $m_{\rm nano}$ is assumed entirely
smoothly distributed in the last panel.
The dotted histogram is the respective panel for
the case with 100\% of the matter in $m_{\rm micro}$.
\label{fig7}}
\end{figure}

\end{document}